\documentclass[aps,superscriptaddress]{revtex4}

\usepackage{amsmath}
\usepackage{amssymb}
\usepackage{bm}
\usepackage{graphicx}
\newcommand{\be}{\begin{equation}}
\newcommand{\ee}{\end{equation}}
\newcommand{\spin}{s}
\begin{document}

\title{Light Cone Matrix Product}
\author{M.~B.~Hastings}
\affiliation{Center for Nonlinear Studies and Theoretical Division, Los Alamos
National Laboratory, Los Alamos, NM 87545}
\begin{abstract}
We show how to combine the light-cone and matrix product algorithms to
simulate quantum systems far from equilibrium for long times.
For the case of the XXZ spin chain at $\Delta=0.5$, we simulate to a
time of $\approx 22.5$.
While part of
the long simulation time is due to the use of the light-cone method,
we also describe a modification of the iTEBD algorithm with improved
numerical stability, and we describe how to incorporate symmetry into
this algorithm.  While statistical
sampling error means that we are not yet able to make a definite statement,
the behavior of the simulation at long times indicates the appearance of
either ``revivals" in the order parameter as predicted by \cite{hl} or
of a distinct shoulder in the decay of the order parameter.
\end{abstract}
\maketitle

Over the last few years, it has become possible to simulate time dynamics of
one dimensional quantum systems for very long times
using matrix product methods such as the
time-evolving bond decimation (TEBD) algorithm\cite{tebd}.   The computational effort required, in both
time and memory, scales exponentially in the entanglement entropy of the
system, and hence these methods work most efficiently
when the quantum state being simulated has a small amount of entanglement.
There are, fortunately, many important examples where the entanglement
grows only logarithmically\cite{us} and these techniques work well.
Broadly speaking, slow entanglement growth seems to happen
when we take a systm in its ground state, and then perturb it
locally, by adding a particle or other perturbation at a single point.

On the other hand, a global perturbation of the system, such as
by starting the system in the ground state of one Hamiltonian and then
evolving it under another, often gives rise to a {\it linear} growth
of entanglement entropy with time, making simulation exponentially
difficult in time.  This behavior has been studied from several
directions, including numerically\cite{numlin}, by conformal field
theory techniques\cite{cftlin}, and by mathematical physics methods
giving general upper bounds on the entanglement growth in time\cite{mp1,mp2}.

One prototypical example of this behavior is the time dynamics of the
spin-$1/2$ XXZ spin chain, 
\be
\label{xxz}
H=\sum_i S^x_i S^x_{i+1}+S^y_i S^y_{i+1} + \Delta S^z_i S^z_{i+1}.
\ee
We start the system at time $t=0$ at the ground state
of the Hamiltonian with infinite $\Delta$, and then evolve the
system for $t>0$ under the Hamiltonian with some finite $\Delta$.
This is a sudden quench from infinite $\Delta$ to finite $\Delta$.
The starting state at infinite $\Delta$ is given by
\be
|\psi\rangle=|...\uparrow\downarrow\uparrow\downarrow...\rangle
\ee
This problem was studied in \cite{numlin} where the linear entropy
growth was found.  A related problem of particles in an optical lattice
was also studied with matrix product methods\cite{proposed}.

In this paper, we present a new approach to simulating such systems far
from equilibrium.  This
approach combines the ``light-cone" method introduced in \cite{lc} with
matrix product techniques.  The result enables us to simulate for significantly
longer times than possible with any other existing method. 

The integrability of the XXZ Hamiltonian does not play any role in our approach.  However, some of the
physical results seen in our simulations may be a result of integrability, as discussed later.  Integrability
has been exploited to study the time dynamics of a BCS pairing model\cite{fcc} after a sudden quench.
The model studied in \cite{fcc} had no spatial structure to the interactions; instead, each fermionic
mode interacted with each other fermionic model, which perhaps makes that model simpler to treat.
However, even for that model it was necessary to use sophisticated numerical calculations to
exploit integrability, so that for the XXZ Hamiltonian above it is not surprising that
we must use numerics to understand the time dynamics.

Our main physical interest in this system is to study the possibility of
``revivals" in the order parameter as predicted by a mean-field study of
the system in \cite{hl}.  Interestingly, the mean-field of \cite{hl} is
the same as the classical limit of the BCS pairing model of \cite{fcc},
although with very different initial conditions.
If one measures the expectation value of
$S^z$ on a given site as a function of time, $\langle S^z_i(t)\rangle$,
one observes an oscillating behavior as a function of time, with damped
oscillations (similar oscillating behavior is also observed for
a related bosonic system in \cite{proposed}).
The expectation value also alternates sign as a function
of site index $i$.  By ``revival", we mean that the envelope of this
damped oscillating function may stop decreasing and instead increase for
short periods of time; overall, the envelope decreases, but for short
periods it may stop decreasing.

It is not clear how applicable the mean-field theory of \cite{hl} should
be to the XXZ chain as
the XXZ
spin chain may be quite far from
mean-field, being in one-dimension.  Further, in this paper
we consider the case $\Delta=0.5$, which is a relatively strong
interaction.  However, we were interested to numerically test the
dynamics of the order parameter at large times to see if some
qualitative features of the mean-field carry over.  Unfortunately,
even at $\Delta=0.5$, the first revivals in the mean-field occur at
a large time ($>20$) which means that existing matrix product algorithms
cannot reach the time to see this behavior.  The light-cone method in
this paper does
reach the desired times.  Due to sampling error in the Monte Carlo sampling,
we are not able to make a definite statement that there are revivals based
on our results, however the results strongly support the possibility of
either revivals (an increase in the envelope) or at least a tendency for
the envelope function to remain constant for periods of time rather than
decreasing.

The next section of the paper explains the basic idea of the light-cone, in a simpler
setting of discrete time evolution.  We show how to significantly reduce the
computational cost involved in computing (on a classical computer) the expectation value of an operator after
applying a quantum circuit to a state; the cost remains exponential, but with a lower
exponential, albeit at the cost of some additional statistical sampling.  After that, we present
the matrix product method we used, a modified version of the infinite time-evolving bond
decimation algorithm (iTEBD)\cite{itebd} with
improved numerical stability.  This section is logically
separate from the rest of the paper; on the one hand, one could use this modified iTEBD on its own, rather
than as part of a light-cone simulation, while on the other hand the rest of the paper simply relies on
using some matrix product algorithm to do the early time simulation and indeed other matrix product algorithms
would work here.  We chose this algorithm since it was best suited to our purposes with the
least numerical effort.  After that section, we describe how to combine the light-cone and quantum circuit
methods, and in the section after that we describe our numerical results.  All numerical work in this
paper is done for the XXZ chain with $\Delta=0.5$.

\section{Light Cone for Quantum Circuits}
The algorithm in this paper is based on the idea of the ``light-cone".  In
relativistic systems, the importance of the light-cone is well-known.
For such systems,
no influence occurs outside the light-cone; equivalently, any
two operators which are space-like separated commute with each other.
However,
even in a system described by non-relativistic quantum mechanics, such
as a one-dimensional spin chain, there is still an
upper limit to the speed at which
any influence can propagate through the system.  This bound is expressed
formally through Lieb-Robinson bounds\cite{lr1,lr2,lr3,lr4}.  Consider any
operator $O$ which acts on a single site, say site number $0$.
The Lieb-Robinson bounds can
be used to show for many systems, including the XXZ spin chain, that
$\exp(-i Ht) O \exp(i H t)\equiv O(t)$ can be approximately described by an
operator
which has support only on a set of sites within distance $v_{LR} t$
of site $0$, where $v_{LR}$ is called the Lieb-Robinson velocity.  In
contrast to relativistic systems, there is some ``leakage" outside this
light-cone; however, we can make this leakage exponentially small by
slightly enlarging the support of the operator we use to approximate
$O(t)$.

In this section, we explain how the presence of a light-cone can be
used to simplify the calculation of time-dependent expectation values.  We explain
this idea in a simpler setting, a quantum circuit model, in order to avoid
the complexities of the Lieb-Robinson bound.  Suppose we have $N$ qubits
on a line, labelled $-N/2,-N/2+1,...,N/2-1$,
and we consider a discrete time dynamics as follows: on the
first time step (and on all subsequent odd time steps),
we act with a set of $2$-qubit gates which act on qubits
$-N/2$ and $-N/2+1$, qubits $-N/2+2$ and $-N/2+3$, and so on.  Then, on the second time step
(and on all subsequent even time steps)
we act with $2$-qubit gates on qubits $-N/2+1$ and $-N/2+2$, qubits $-N/2+2$ and $-N/2+3$ and so on.

We consider the following problem, which is a discrete-time analogue
of the continuous time problem addressed elsewhere in this paper.
We initialize the system to some given product state, evolve for $T=N/2$ time steps,
and then measure the expectation value of the $z$-component of the
spin at site $0$.

How long does it take to compute this expectation value on a classical
computer?  The simplest algorithm is to store the amplitudes for the quantum
state as a $2^N$-dimensional complex vector, and update this amplitude at
each time step.  The time required for a single time step for this algorithm
is of order
$2^N=2^{2T}$, and the total time is of order $T 2^{2T}$.

We now explain how to reduce this exponential from $2^{2T}$ to $2^T$, at
the cost of having to do some statistical sampling and of only
approximating the expectation value.  Consider Fig.~1, which shows
a drawing of the gates in time.  First, note that the gates outside the
triangle
have no effect on the final output, lying outside the
discrete light-cone, and so they can be ignored.

Before explaining how to actually do the calculation, let us motivate
the approach physically, using the idea of entanglement.  After time $T/2$,
it is only necessary to consider the dynamics within region $A$ marked in
Fig.~1.  Since this dynamics occurs only on the sites within
distance $T/2=N/4$ of site $0$, it occurs on a system of length $T=N/2$,
and hence the cost
to simulate {\it pure state} evolution on these sites is of order $2^{T}$.
On the other hand, at early times
(before time $T/2$)
the system has less entanglement, and so also should be easier to simulate.
The difficulty with implementing this physical idea is that at time $T/2$,
the reduced density matrix on sites $-T/2,...,T/2$ is not a pure state
as those sites are entangled with the region outside.  Simulating
the time dynamics of a mixed state is much more costly than simulating a pure state, taking
$2^{2T}$ rather than $2^T$.  However, imagine
that someone did a projective measurement of the system in some arbitrary
basis of states on sites $-N/2,...,-N/4-1$
at time $T/2$, and someone else did a projective measurement on sites
$N/4+1,...,N/2-1$.  The measurement has no effect on the expectation value of
the spin on site $0$.  However, conditioned on the outcome of the measurement,
the reduced density matrix in $-T/2,...,T/2$ becomes a pure state.
Thus, we can statistically sample different measurement outcomes and
then do a simulation of pure state dynamics in $-T/2,...,T/2$.

Now we explain the detailed approach.  Let the initial state $\Psi_0=\Psi_L
\otimes \Psi_R$, where $\Psi_{L,R}$ are states on the left and right half
of the system.  Let $U_A,U_B,U_C,U_D$ be the unitary operators associated
with the gates in regions $A,B,C,D$.  Let $\Pi^L_\alpha$ 
denote
projection operators onto a complete basis of states, labelled by index
$\alpha$, on sites $-N/2,...,-N/4-1$ and let $\Pi^R_\beta$
denote
projection operators onto a complete basis of states, labelled by index
$\beta$, on sites $N/4+1,...,N/2$.
We wish to compute
\begin{eqnarray}
&& 
\langle \Psi_L\otimes \Psi_R|U_B^\dagger U_D^\dagger U_C^\dagger U_A^\dagger
S^z_0 U_A U_C U_D U_B|\Psi_L \otimes \Psi_R \rangle
\\ \nonumber
&=&
\langle (U_B \Psi_L)\otimes (U_D \Psi_R)|U_C^\dagger U_A^\dagger
S^z_0 U_A U_C| 
(U_B \Psi_L)\otimes (U_D \Psi_R) \rangle
\\ \nonumber
&=&
\sum_{\alpha,\beta}
\langle (U_B \Psi_L)\otimes (U_D \Psi_R)|
\Pi^L_\alpha \Pi^R_\beta
U_C^\dagger U_A^\dagger
S^z_0  U_A U_C
\Pi^L_\alpha \Pi^R_\beta
|(U_B \Psi_L)\otimes (U_D \Psi_R) \rangle
\\ \nonumber
&=&
\sum_{\alpha,\beta}
\langle (U_B \Psi_L)\otimes (U_D \Psi_R)|
\Pi^L_\alpha \Pi^R_\beta
|
(U_B \Psi_L)\otimes (U_D \Psi_R) \rangle
\\ \nonumber
&&\times
\frac{
\langle (U_B \Psi_L)\otimes (U_D \Psi_R)|
\Pi^L_\alpha \Pi^R_\beta
U_C^\dagger U_A^\dagger
S^z_0  U_A U_C
\Pi^L_\alpha \Pi^R_\beta
|
(U_B \Psi_L)\otimes (U_D \Psi_R) \rangle}
{
\langle (U_B \Psi_L)\otimes (U_D \Psi_R)|
\Pi^L_\alpha \Pi^R_\beta
|
(U_B \Psi_L)\otimes (U_D \Psi_R) \rangle},
\end{eqnarray}
where the second equality follows because the projection operators $\Pi^{L}_\alpha,\Pi^R_\beta$
commute with the unitaries $U_A,U_C$; i.e., because the projection operators
are outside the light-cone.
Interpreting
\be
\label{weightqc}
\langle (U_B \Psi_L)\otimes (U_D \Psi_R)|
\Pi^L_\alpha \Pi^R_\beta
|
(U_B \Psi_L)\otimes (U_D \Psi_R) \rangle
\ee
as a statistical weight, we can compute the desired result by sampling
\be
\label{tosamp}
\frac{
\langle (U_B \Psi_L)\otimes (U_D \Psi_R)|
\Pi^L_\alpha \Pi^R_\beta
U_C^\dagger U_A^\dagger
S^z_0  U_A U_C
\Pi^L_\alpha \Pi^R_\beta
|
(U_B \Psi_L)\otimes (U_D \Psi_R) \rangle}
{
\langle (U_B \Psi_L)\otimes (U_D \Psi_R)|
\Pi^L_\alpha \Pi^R_\beta
|
(U_B \Psi_L)\otimes (U_D \Psi_R) \rangle}
\ee
with weight (\ref{weightqc}).
Define $\Psi_{\alpha,\beta}$ by
\be
|\Psi_{\alpha,\beta}\rangle=
\frac{\Pi^L_{\alpha} \Pi^R_{\beta}|(U_B \Psi_L)\otimes (U_D \Psi_R) \rangle}
{\Bigl|\Pi^L_{\alpha} \Pi^R_{\beta}|(U_B \Psi_L)\otimes (U_D \Psi_R) \rangle\Bigr|}.
\ee
Then Eq.~(\ref{tosamp}) is equal to
\be
\langle \Psi_{\alpha,\beta}|S^z_0
|\Psi_{\alpha,\beta}\rangle,
\ee
and
$\langle \Psi_L\otimes \Psi_R|U_B^\dagger U_D^\dagger U_C^\dagger U_A^\dagger
S^z_0 U_A U_C U_D U_B|\Psi_L \otimes \Psi_R \rangle$ is equal to
\be
\overline{\langle \Psi_{\alpha,\beta}|S^z_0
|\Psi_{\alpha,\beta}\rangle},
\ee
where $\overline{\langle ... \rangle}$ denotes an average over Monte Carlo steps with
weight (\ref{weightqc}).

The light-cone algorithm for this quantum circuit is to sample (\ref{tosamp}) with
weight (\ref{weightqc}).
All the calculations described here, such as calculating
$U_B \Psi_L$ or 
$U_D \Psi_R$, can be done in a time of order  $T \exp(T)$,
rather than $T \exp(2T)$, and hence the light-cone algorithm takes
a time of order $T\exp(T)$ for each Monte Carlo sample.  Since the operator
$S^z_0$ has bounded operator norm, the root-mean square fluctuations in the
expectation value are bounded, and hence the sampling error decreases as
one over the square-root of the number of samples.

\begin{figure}
\centerline{
\includegraphics[scale=0.7]{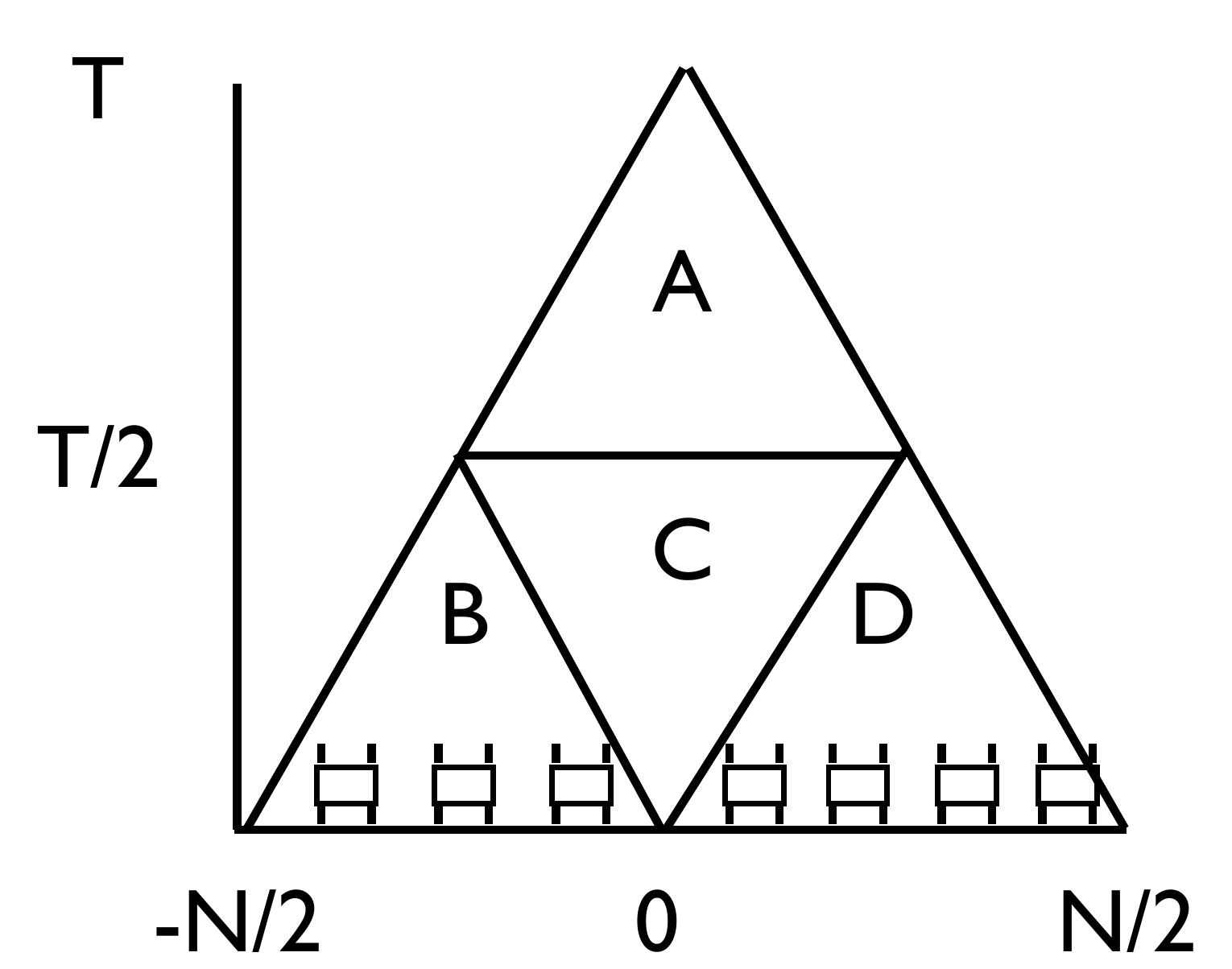}}
\caption{Using the light-cone method for a quantum circuit.  Triangle
includes sites within the light-cone.  Qubits are arranged along the
horizontal axis, time increases vertically.  We show the first round of the
quantum circuit.
Regions A,B,C,D are as described in the text.}
\vspace{5mm}
\end{figure}

The basic idea of the light-cone quantum circuit method
described in this section is the same idea used in \cite{lc} where it
was applied to continuous time dynamics.  In the rest of this
paper, we again treat continuous time dynamics, but in contrast
to \cite{lc} we use a matrix
product algorithm to perform the early time simulation.  The use of
the matrix product algorithm
relies on the limited entanglement at early times
(this contrasts with the approach in this section,
where instead we factorized the early time dynamics into a product of two different unitaries,
$U_B$ and $U_D$, and relied on the product nature of the initial state).  One complexity in the matrix product
simulation is that the probability distribution of outcomes $\alpha,\beta$
will not factorize into a product of separate distributions for $\alpha$
and $\beta$ because there may be some entanglement between those portions
of the system.  We will explain how to resolve this problem.  Another difference
in the calculations in the rest of the paper is that the early time evolution
using the matrix product evolution will be done for a time which is slightly
more than half the final time, rather than exactly half as it is here.

\section{Matrix Product Algorithm}
\subsection{Modified iTEBD}
To do the matrix product simulations, we used
a modified version of the iTEBD algorithm with improved
numerical stability.  Before describing the improved algorithm, let us
recall how iTEBD works.  The algorithm stores a matrix product state
for an infinite system.  We have matrices
$\Gamma^A(s)$ and $\Gamma^B(s)$, for the even and odd sites
respectively, where
$s$ labels the spin on the site.  There are also diagonal matrices
$\lambda^A$ and $\lambda^B$ defined on the bonds directly to the right
of the even and odd sites.
The wavefunction amplitude for a given configuration of
spins, $...,s_0,s_1,s_2,...$ is equal to
$\psi(....,s_0,s_1,s_2,....)=...\Gamma^A(s_0)
\lambda^A \Gamma^B(s_1) \lambda^B \Gamma^A(s_2) \lambda^A...$
The matrices $\lambda^{A,B}$ are chosen such that their diagonal coefficients
coincide with the Schmidt coefficients of the state decomposed
across the given bond; this contrasts with other matrix product state
representations used in other algorithms where the algorithm only
stores a single matrix for each site.

The iTEBD matrix product state can represent a system with translational
invariance of period 2.  It is thus very convenient to use for our system,
which has such invariance in its initial conditions.  The time evolution
is simulated using a Trotter-Suzuki method.  The time evolution is
broken up into small steps $\delta t$, and 
the time evolution is approximated by
a series of two-site unitary transformations,
$U^{AB}=\otimes_{r} U^{[2r,2r+1]}$ and $U^{BA}=\otimes_{r} U^{[2r-1,2r]}$,
where $U^{[r,r+1]}\equiv \exp(-i h^{[r,r+1]} \delta t)$
is a unitary acting on sites $r,r+1$.  By updating first
with $U^{AB}$ and then with $U^{BA}$, the state maintains translational
invariance at all times.  In our implementation of the
modified iTEBD algorithm
described below
we use a second-order Trotter-Suzuki decomposition with a time step of
$0.0625$.

In a single update step, the iTEBD algorithm proceeds as follows.  Consider
the update under $U^{AB}$.  The matrix $U^{[2r,2r+1]}$ has matrix
elements
$U^{[2r,2r+1]}_{s_A,s_B;s_A',s_B'}$, where $s_A,s_B$ represent
the spins on the $A$ and $B$ sites after acting with $U$ and $s_A',s_B'$
represent the spins on the $A$ and $B$ sites before acting with $U$.
First, compute the matrix
\be
\label{Thetadef}
\Theta(s_A,s_B)\equiv \lambda^B C(s_A,s_B),
\ee
where we define the matrix $C$ by
\be
\label{Cdef}
C(s_A,s_B)=\sum_{s_A',s_B'} 
U^{[2r,2r+1]}_{s_A,s_B;s_A',s_B'}
\Gamma^A(s_A') \lambda^A \Gamma^B(s_A') \lambda^B.
\ee
Thus,
\be
\label{Thetafull}
\Theta(s_A,s_B)=
\sum_{s_A',s_B'} 
U^{[2r,2r+1]}_{s_A,s_B;s_A',s_B'}
\lambda^B 
\Gamma^A(s_A') \lambda^A \Gamma^B(s_A') \lambda^B.
\ee
This matrix $C$ is not part of the usual description of the iTEBD algorithm,
but we define it for use later.  Note that in Eq.~(\ref{Cdef}) the
quantity
$U^{[2r,2r+1]}_{s_A,s_B;s_A',s_B'}$ is a complex number, and all
other quantities $\Gamma^{A,B}$ and $\lambda^{A,B}$ represent
matrices which are multiplied following the usual
rules of matrix multiplication.

The matrix $\theta(s_A,s_B)$ has matrix
elements
$\theta(s_A,s_B)_{\alpha,\beta}$.  Thus,
$\theta$ is a {\it four} index tensor, with indices $s_A,s_B,\alpha,\beta$.
In the second step, do a singular value decomposition of $\theta$
according to the index bipartition $\alpha,s_A$ and $\beta,s_B$, so
that
\be
\label{svd}
\theta(s_A,s_B)_{\alpha,\beta}
=\sum_\beta X_{\alpha,s_A;\beta} \tilde \lambda_{\beta\beta}^A Y_{\beta;\gamma,s_B}.
\ee
The subscript notation $\alpha,s;\beta$ is intended to indicate that
we treat $X$ as a matrix with rows labelled by $\alpha$ and $s$ and
columns labelled by $\beta$.

In the third step,
define
the matrices $\tilde \Gamma^{A,B}$ by their matrix elements
$\tilde \Gamma^{A,B}(s)_{\alpha\beta}$ as follows:
\begin{eqnarray}
\label{tgdef}
\tilde \Gamma^A(s)_{\alpha\beta}=\Bigl(\lambda^B_{\alpha\alpha}\Bigr)^{-1} X_{\alpha,s;\beta}, \\ \nonumber
\tilde \Gamma^B(s)_{\beta\gamma}=Y_{\beta;\gamma,s} \Bigl(\lambda^B_{\gamma\gamma}\Bigr)^{-1}.
\end{eqnarray}
The matrices $\tilde \Gamma^A,\tilde \Gamma^B,\tilde \lambda^A$ replace
the old matrices $\Gamma^A,\Gamma^B,\lambda^A$ after this update step,
while the matrix $\lambda^B$ remains unchanged under the update step.
Finally, we perform a truncation of the state: after each step, the
bond dimension increases, and we truncate the state by keeping only
the $k_{max}$ largest singular values of $\theta$, discarding the others.
The quantity $k_{max}$ determines the maximum bond dimension, with
increasing $k_{max}$ leading to an increased accuracy in return for
additional numerical effort.  The step with the largest numerical cost is the singular
value decomposition, which requires a computational time scaling as ${\cal O}(k_{max}^3)$.

Unfortunately, we found some difficulties with numerical stability in our
implementation of this algorithm for larger $(>1000)$ values of $k_{max}$.
The matrices $\lambda^{A}$ will contain some very small diagonal
entries, and therefore $(\lambda^{A})^{-1}$ will be very large.  Therefore,
any small error in the singular value decomposition will tend to get
magnified when multiplying by $(\lambda^{A})^{-1}$.  Note the sequence of
steps: first we multiply by $\lambda^{A}$, then we do a singular value
decomposition, and then we multiply by $(\lambda^{A})^{-1}$, so that
problems can arise if the singular value decomposition in between the
multiplication steps is not numerically exact.

The solution to this problem is simple.  First, we changed which matrices are
stored by the algorithm, so that instead of storing the matrices
$\Gamma^A,\Gamma^B$, we store the matrices $A^A,A^B$ defined by
\begin{eqnarray}
\label{Adef}
A^A(s)\equiv \Gamma^A(s) \lambda^A, \\ \nonumber
A^B(s)\equiv \Gamma^B(s) \lambda^B.
\end{eqnarray}
These matrices $A^A,A^B$ are the usual $A$ matrices defining a matrix product state.
Such a relation $A(s)=\Gamma(s)\lambda$ was used previously in \cite{long}, but
\cite{long} used a method
of doing the update which still required a division by $\lambda$ while the
point of the method described below is that it does not require any division
(other minor differences are that \cite{long} was not in the context
of infinite systems and also in \cite{long}
reduced density matrices
were diagonalized while
we employ singular value decomposition instead).

We continue to store the matrices $\lambda^{A},\lambda^{B}$.
Then, in an update step, we first compute
\be
\label{NewThetaCompute}
C(s_A,s_B)\equiv
\sum_{s_A',s_B'} 
U^{[2r,2r+1]}_{s_A,s_B;s_A',s_B'}
A^A(s_A') A^B(s_A').
\ee
Note that if we substitute
Eq.~(\ref{Adef}) into
(\ref{NewThetaCompute}) we find the same result for $C$ as (\ref{Cdef}).
We then compute the matrix $\Theta(s_A,s_B)$ from $C(s_A,s_B)$ following
Eq.~(\ref{Cdef}).
We next do a singular value decomposition of
$\theta$ as in Eq.~(\ref{svd}), obtaining matrices $X,Y,\tilde\lambda^A$.
In analogy to Eq.~(\ref{Adef}), we now wish to compute
the updated
matrices $\tilde A$ which are defined by:
\begin{eqnarray}
\label{tAdef}
\tilde A^A(s)\equiv \tilde \Gamma^A(s) \lambda^A, \\ \nonumber
\tilde A^B(s)\equiv \tilde \Gamma^B(s) \lambda^B.
\end{eqnarray}

However, in order to compute the updated matrices $\tilde A^{A},\tilde A^{B}$,
we do not do this by computing $\tilde \Gamma^A,\tilde \Gamma^B$ and applying (\ref{tAdef}).
Instead, we
compute them in a different way.  To derive the approach that we use, note that
Eq.~(\ref{tgdef}) implies that Eq.~(\ref{tAdef}) is equivalent to
\be
\label{see1}
\tilde A^B(s)_{\beta,\gamma}=Y_{\beta;s,\gamma}
\ee
and 
\begin{eqnarray}
\label{see2}
\tilde A^A(s_A)_{\alpha\beta}&=&\Bigl(\lambda^B_{\alpha\alpha}\Bigr)^{-1} X_{\alpha,s;\beta}
\\ \nonumber
&=& 
\Bigl(\lambda^B_{\alpha\alpha}\Bigr)^{-1} 
\sum_{\gamma,s_B}
\Theta(s_A,s_B)_{\alpha\gamma} 
\Bigl(Y^{-1}\Bigr){\gamma,s_B;\beta}.
\\ \nonumber
&=& 
\sum_{\gamma,s_B}
C(s_A,s_B)_{\alpha\gamma} 
\Bigl(Y^{-1}\Bigr){\gamma,s_B;\beta},
\end{eqnarray}
where the matrix $Y^{-1}$ has the opposite index bipartition to matrix $Y$ (that is, $\gamma,s_B;\beta$ instead of
$\beta;\gamma,s_B$).

Note that Eq.~(\ref{see1}) does not involve multiplying by
$\lambda^{-1}$.  Similarly, the last line of Eq.~(\ref{see2}) gives an
expression for $\tilde A^a$ that does not involve multiplying by $\lambda^{-1}$,
so this method of computing $\tilde A$ does not have the same problem
of numerical stability.  Further, no extra CPU time is required to
compute $C$ on the last line of (\ref{see2})
since the algorithm computes this
already
as part of computing the matrix $\theta$.  Finally, since $Y$ is unitary,
we have $Y^{-1}=Y^\dagger$ so there is no overhead to compute the inverse of
$Y$.
That is, Eq.~(\ref{see2}) implies that
\be
\label{see3}
\tilde A^A(s_A)_{\alpha\beta}=
\sum_{\gamma,s_B}
C(s_A,s_B)_{\alpha\gamma} 
Y^*_{\beta;\gamma,s_B}.
\ee

Thus, we can use Eq.~(\ref{see1},\ref{see3}) to improve the stability of the
algorithm.
The only extra overhead required
is one extra matrix multiplication, to multiply by $Y^{\dagger}$ as in (\ref{see3}).
The overhead to do this is small compared to the time required to do the
singular value decomposition of $\theta$.
The complete pseudo-code for a single update step is:
\begin{itemize}
\item[{\bf 1.}] Compute $C$ by Eq.~(\ref{NewThetaCompute}) and then compute $\Theta$ by Eq.~(\ref{Thetadef}).

\item[{\bf 2.}] Do a singular value decomposition as in Eq.~(\ref{svd}).

\item[{\bf 3.}] Update $\tilde A^A,\tilde A^B$ by Eqs.~(\ref{see1},\ref{see3}).
\end{itemize}

\subsection{Incorporating Symmetry}
Both the iTEBD algorithm and the modified iTEBD algorithm discussed here make it
very easy to incorporate symmetry such as conservation of total $S^z$ as follows.
For a Hamiltonian such as the XXZ Hamiltonian (\ref{xxz}) and for initial
conditions which have definite $S^z$ on each site as we have considered, it
is always possible to choose the basis vectors of the Schmidt
decomposition to have definite $S^z$.
Thus, for each bond variable $\alpha$, we can assign a definite $S^z$, giving the
total $S^z$ of the spins to the left site of the given bond.  The infinite
size of the system does not give any trouble with assigning a definite $S^z$,
as we simply define the $S^z$ of the system to the left of a given bond to
be the {\it difference} between the total spin to the left and the initial
spin to the left.  This difference can be easily kept track of because it
increases or decreases
by one when a single spin up moves across the bond to the left or
right.

After doing the singular value decomposition for each $S^z$, we get
a list of Schmidt coefficients for each $S^z$.  If there are more than
$k_{max}$ different Schmidt coefficients, we truncate.  To do the truncation,
we merge these lists, sort the merged list from largest to smallest,
find the $k_{max}$ largest Schmidt coefficients, and  keep those.  This
means that the number of Schmidt coefficients we keep for each $S^z$
will depend on $S^z$; we find that we keep many Schmidt coefficients
for $S^z$ near zero, and fewer for $S^z$ far from zero.

Some fluctuation is observed in how many Schmidt coefficients were kept for each $S^z$, but
a typical set of values for $k_{max}=4096$ was $2$ at $S^z=-5$; $28$ at $S^z=-4$; $142$ at $S^z=-3$; $403$ at
$S^z=-2$; $748$ at $S^z=-1$; $968$ at $S^z=0$; $890$ at $S^z=+1$; $576$ at $S^z=+2$; $254$ at $S^z=+3$; $71$ at
$S^z=+4$; $12$ at $S^z=+5$; and $2$ at $S^z=+6$.  Note the asymmetry between positive and negative $S^z$, since
the alternating spins in the initial configuration, combined with our definition of $S^z$ as the total spin
to the left of a given bond, breaks the symmetry.  To explain this asymmetry further, consider a site which
in the initial configuration is spin up, with its neighbor spin down; the only way that the spin to the left
of this bond can change at the first instant of time is for the up spin on the given site to hop to the left,
increasing the total spin to the left, but there is no corresponding process which would decrease the
spin to the left in the first instant of time.

The most computationally costly step is singular value decomposition.  The bond variable $\beta$ in
Eq.~(\ref{svd}) has an $S^z$ which is equal to the $S^z$ of bond variable $\alpha$ plus the $z$-spin of
$s_A$.  Thus since $s_A$ takes two different values,
there are two different $S^z$ values of bond variables $\alpha$ which contribute to a given $S^z$ value for
bond variable $\beta$.
Thus, for the specific number of Schmidt coefficients kept which is given in the above paragraph, the largest
matrix that we decompose has bond dimension
$968+890=1858$.  Without the use of symmetry, keeping $4096$ bond coefficients would require decomposing matrices
of size $8192$.

\section{Combining Light-Cone and Matrix Product}
We now describe how we combine the matrix product algorithm with the light-cone method.
The matrix product simulation will be accurate for a certain range of times; the larger
$k_{max}$ is the longer it works.  We run the algorithm for as long a time as possible
for the given $k_{max}$.  Let this time be $t_{init}$.  For $k_{max}=4096$, we could take
$t_{init}$ slightly larger than $16$ before encountering appreciable truncation error.
The $k_{max}$ required to reach a given $t_{init}$ at a given truncation error
increased exponentially with $t_{init}$.
After simulating to time $t_{init}$,
we save the matrices defining the matrix product state.  A separate program then
performs the second phase of the simulation as follows.

We wish to compute the expectation value of the $z$-component of the
spin on a site, say site $0$, at a time $t_{fin}>t_{init}$.  That is, $\langle S^z_0(t_{fin}) \rangle$.  The matrix
product simulation gives a matrix product state $|\psi^{mps}(t_{init})\rangle$ which
is a good
approximation to the state $|\psi(t_{init})\rangle$, where
\be
|\psi(t_{init})\rangle \equiv \exp(-i H t_{init}) |\psi\rangle.
\ee
Then
\be
\label{goal}
\langle S^z_0(t_{fin}) \rangle \approx \langle \psi^{mps}(t_{init})| 
\exp[i H (t_{fin}-t_{init})]
S^z_0 
\exp[-i H (t_{fin}-t_{init})] 
| \psi^{mps}(t_{init})\rangle.
\ee
Even in this continuous time setting, the situation is similar to that
in Fig.~1:
the dynamics of the system outside
this light-cone has little effect on the spin.  The Lieb-Robinson bounds\cite{lr1,lr2,lr3,lr4}
make this intuition precise; using these bounds we can prove that we can approximate (\ref{goal})
by
\be
\label{loc}
\langle S^z_0(t) \rangle \approx \langle \psi^{mps}(t_{init})| 
\exp[i H^{loc} (t_{fin}-t_{init})]
S^z_0 
\exp[-i H^{loc} (t_{fin}-t_{init})] 
| \psi^{mps}(t_{init})\rangle,
\ee
where $H^{loc}$ includes only the interaction of sites within
some distance $l$ of site $i$.  Thus,
\be
H^{loc}=\sum_{i=-l}^{l-1} S^x_i S^x_{i+1}+S^y_i S^y_{i+1} + \Delta S^z_i S^z_{i+1}.
\ee
As long as $l>v_{LR} (t_{fin}-t_{init})$, then it is possible to use
the Lieb-Robinson bounds to prove that the approximation
(\ref{loc}) is exponentially
good, where $v_{LR}$ is the Lieb-Robinson velocity.  Previous
experience with other light-cone techniques tells us that in fact the
approximation is good so long as $l>v_{sw} (t_{fin}-t_{init})$ where $v_{sw}$ is
the spin-wave velocity, given by
\be
v_{sw}=(\pi/2) \sin(\theta)/\theta,
\ee
where $\cos(\theta)=\Delta$\cite{swvel}.  Note that the dynamics we consider
takes place in a highly excited state, rather than in the ground state, so
{\it a priori} it is not obvious that $v_{sw}$ is the correct velocity
for excitations, but numerical evidence both previously and in this work
indicates that it still is the correct velocity in this regime.  Note
also that $v_{sw}<v_{LR}$ for this system.

\begin{figure}
\label{LRf}
\centerline{
\includegraphics[scale=0.7]{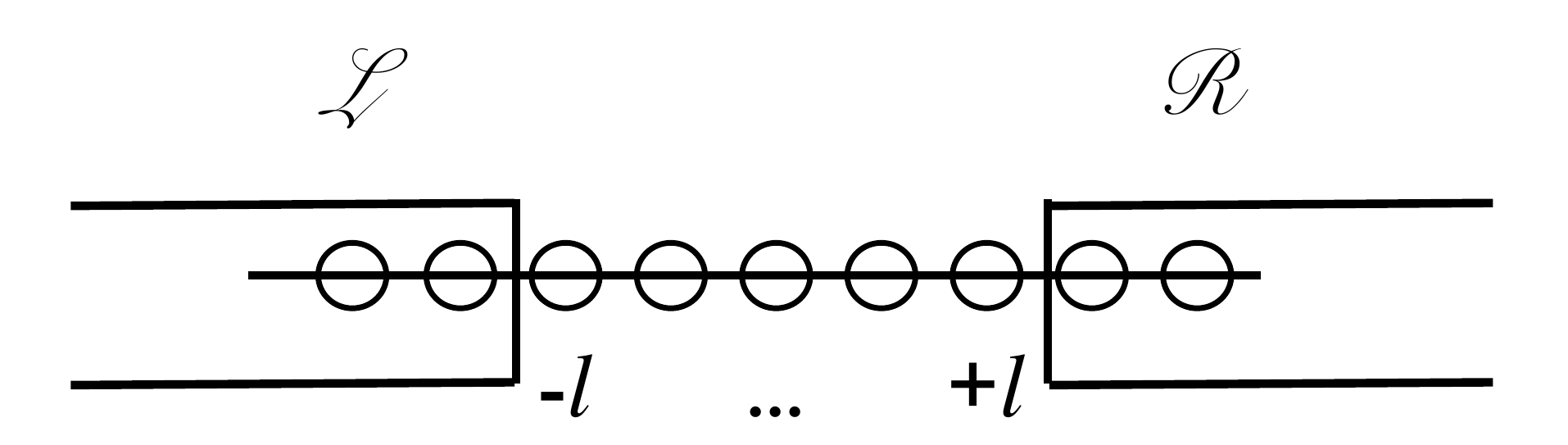}}
\caption{Sketch of regions ${\cal L},{\cal R}$ for $l=2$.}
\vspace{5mm}
\end{figure}

Let ${\cal L}$ denote the set of sites to the left of
site $-l$ and ${\cal R}$ denote the set of sites to the
right of site $+l$, as shown in Fig.~2.
Imagine that at time $t_{init}$ someone measured that state of the system on $L$, doing
the measurement in the Schmidt basis, and also measured the state of
the system on $R$, again in the Schmidt basis.  Physically, this
measurement, being outside the light-cone,
should not affect the measurement of the spin on site
$0$ at time $t_{fin}$.  Mathematically, we can express this as
\begin{eqnarray}
\label{correct}
&&\langle \psi^{mps}(t_{init})| 
\exp[i H^{loc} (t_{fin}-t_{init})]
S^z_0 
\exp[-i H^{loc} (t_{fin}-t_{init})] 
| \psi^{mps}(t_{init})\rangle \\ \nonumber
&=&
\sum_{\alpha,\beta} 
\langle \psi^{mps}(t_{init})| 
\Pi^L_{\alpha}
\Pi^R_{\beta}
\exp[i H^{loc} (t_{fin}-t_{init})]
S^z_0 
\exp[-i H^{loc} (t_{fin}-t_{init})] 
\Pi^L_{\alpha}
\Pi^R_{\beta}
| \psi^{mps}(t_{init})\rangle,
\end{eqnarray}
where $\Pi^L_{\alpha}$ projects onto state $\alpha$ in $L$ and
$\Pi^R_{\beta}$ projects onto state $\beta$ in $R$, with
$\sum_{\alpha} \Pi^L_{\alpha}=\openone$.  Note that the
correctness of (\ref{correct}) depends on the fact that
$\Pi^L_{\alpha}$ and $\Pi^R_{\beta}$ commute with
$\exp[i H^{loc} (t_{fin}-t_{init})]
S^z_0 
\exp[-i H^{loc} (t_{fin}-t_{init})]$.

Note that
\begin{eqnarray}
&&
\sum_{\alpha,\beta} 
\langle \psi^{mps}(t_{init})| 
\Pi^L_{\alpha}
\Pi^R_{\beta}
\exp[i H^{loc} (t_{fin}-t_{init})]
S^z_0 
\exp[-i H^{loc} (t_{fin}-t_{init})] 
\Pi^L_{\alpha}
\Pi^R_{\beta}
| \psi^{mps}(t_{init})\rangle
\\ \nonumber
&=& 
\sum_{\alpha,\beta} 
\langle \psi^{mps}(t_{init})| 
\Pi^L_{\alpha}
\Pi^R_{\beta}
| \psi^{mps}(t_{init})\rangle \times
\\ \nonumber
&& 
\frac{
\sum_{\alpha,\beta} 
\langle \psi^{mps}(t_{init})| 
\Pi^L_{\alpha}
\Pi^R_{\beta}
\exp[i H^{loc} (t_{fin}-t_{init})]
S^z_0 
\exp[-i H^{loc} (t_{fin}-t_{init})] 
\Pi^L_{\alpha}
\Pi^R_{\beta}
| \psi^{mps}(t_{init})\rangle}
{\sum_{\alpha,\beta} 
\langle \psi^{mps}(t_{init})| 
\Pi^L_{\alpha}
\Pi^R_{\beta}
| \psi^{mps}(t_{init})\rangle}.
\end{eqnarray}
Thus, if we statistically sample with the weight
\be
\label{weight}
\langle \psi^{mps}(t_{init})| 
\Pi^L_{\alpha}
\Pi^R_{\beta}
| \psi^{mps}(t_{init})\rangle
\ee
we find that
\begin{eqnarray}
\label{s}
&&\langle \psi^{mps}(t_{init})| 
\exp[i H^{loc} (t_{fin}-t_{init})]
S^z_0 
\exp[-i H^{loc} (t_{fin}-t_{init})] 
| \psi^{mps}(t_{init})\rangle \\ \nonumber
\\ \nonumber
&=& 
\overline
{\frac{
\langle \psi^{mps}(t_{init})| 
\Pi^L_{\alpha}
\Pi^R_{\beta}
\exp[i H^{loc} (t_{fin}-t_{init})]
S^z_0 
\exp[-i H^{loc} (t_{fin}-t_{init})] 
\Pi^L_{\alpha}
\Pi^R_{\beta}
| \psi^{mps}(t_{init})\rangle}
{\sum_{\alpha,\beta} 
\langle \psi^{mps}(t_{init})| 
\Pi^L_{\alpha}
\Pi^R_{\beta}
| \psi^{mps}(t_{init})\rangle}}
\\ \nonumber
\\ \nonumber
&=&
\overline
{\langle \psi^{mps}_{\alpha,\beta}(t_{init})|
\exp[i H^{loc} (t_{fin}-t_{init})]
S^z_0 
\exp[-i H^{loc} (t_{fin}-t_{init})] |
\psi^{mps}_{\alpha,\beta}(t_{init})\rangle}
\end{eqnarray}
where $\overline{\langle...\rangle}$ denotes the average over Monte Carlo steps
with weight (\ref{weight}),
and
\be
|\psi^{mps}_{\alpha,\beta}(t_{init})\rangle=
\frac{\Pi^L_{\alpha}\Pi^R_{\beta} |\psi^{mps}(t_{init})\rangle}
{\Bigl|\Pi^L_{\alpha}\Pi^R_{\beta} |\psi^{mps}(t_{init})\rangle\Bigr|}.
\ee

The light-cone algorithm does the statistical sampling of (\ref{s}) with
weight (\ref{weight}).
For each Monte Carlo step, the algorithm does three things. {\bf (1)}: Pick
$\alpha,\beta$ according to the probability distribution (\ref{weight});
{\bf (2)}: Calculate the state
$|\psi^{mps}_{\alpha,\beta}(t_{init})\rangle$ in the computational basis (a product basis in which
each site has definite spin up or down); {\bf (3)}: Evolve the state 
$|\psi^{mps}_{\alpha,\beta}(t_{init})\rangle$ forward in time for a time $t_{fin}-t_{init}$ and compute
the expectation value of $S^z_0$.  We explain how to do each of these steps in turn.

\subsection{Sampling $\alpha,\beta$}
To sample $\alpha,\beta$ according to (\ref{weight}), we proceed in a series of steps.
The difficulty that we encounter is that $\alpha$ and $\beta$ are correlated: the probability
distribution in (\ref{weight}) does not factorize.  We will show how to overcome this
by replacing this probability distribution by a product of conditional probability distributions
which are easier to compute, and then sampling each of those probability distributions.  Finally,
after explaining our approach, we will explain why an alternate approach, which at first
sight seems more natural, is not good because it requires much more computer time.

Note that
\be
\label{cond}
\langle \psi^{mps}(t_{init})| 
\Pi^L_{\alpha}
\Pi^R_{\beta}
| \psi^{mps}(t_{init})\rangle=P(\alpha) P(\beta|\alpha),
\ee
where
\be
P(\alpha)=\langle \psi^{mps}(t_{init})| 
\Pi^L_{\alpha}
| \psi^{mps}(t_{init})\rangle
\ee
and
\be
P(\beta|\alpha)=
\frac{
\langle \psi^{mps}(t_{init})| 
\Pi^L_{\alpha}
\Pi^R_{\beta}
| \psi^{mps}(t_{init})\rangle}
{
\langle \psi^{mps}(t_{init})| 
\Pi^L_{\alpha}
| \psi^{mps}(t_{init})\rangle
}.
\ee
Note that
\be
\sum_{\beta}
\frac{
\langle \psi^{mps}(t_{init})| 
\Pi^L_{\alpha}
\Pi^R_{\beta}
| \psi^{mps}(t_{init})\rangle}
{
\langle \psi^{mps}(t_{init})| 
\Pi^L_{\alpha}
| \psi^{mps}(t_{init})\rangle
}
=1,
\ee
so that $P(\beta|\alpha)$ can indeed be interpreted as a conditional probability distribution.
Therefore, we can first pick $\alpha$ randomly
according to the probability distribution
$P(\alpha)$
and then choose $\beta$ randomly according to the probability distribution $P(\beta|\alpha)$.
Choosing $\alpha$ is easy, since $P(\alpha)$ is given by the square of the corresponding
Schmidt coefficient, and these Schmidt coefficients are the diagonal entries of $\lambda$.
Choosing $\beta$ is more difficult, since $\alpha$ and $\beta$ are correlated.

To choose $\beta$, we proceed as follows.  Define $\Pi^i_{\uparrow}$ to project onto
the state of site $i$ with spin up, and define $\Pi^i_{\downarrow}$ to project onto
the state with spin down.  Then, we have
\be
P(\beta|\alpha)=\sum_{\spin_{-l}=\uparrow,\downarrow} P(\spin_{-l}|\alpha) P(\beta|\spin_{-l}\alpha)
,
\ee
where
\be
P(\spin_{-l}|\alpha)=
\frac{
\langle \psi^{mps}(t_{init})| 
\Pi^{-l}_{\spin}
\Pi^L_{\alpha}
| \psi^{mps}(t_{init})\rangle}
{
\langle \psi^{mps}(t_{init})| 
\Pi^L_{\alpha}
| \psi^{mps}(t_{init})\rangle},
\ee
and
\be
P(\beta|\spin_{-l}\alpha)=
\frac{
\langle \psi^{mps}(t_{init})| 
\Pi^R_{\beta}
\Pi^{-l}_{\spin}
\Pi^L_{\alpha}
| \psi^{mps}(t_{init})\rangle}
{
\langle \psi^{mps}(t_{init})| 
\Pi^L_{\alpha}
\Pi^{-l}_{\spin}
| \psi^{mps}(t_{init})\rangle}.
\ee
Note that
$P(\uparrow|\alpha)+
P(\downarrow|\alpha)=1$ so that we can interpret $P(\spin|\alpha)$ as a conditional
probability.

Repeating this, we find that
\be
P(\beta|\alpha)=\sum_{\spin_l} \Bigl( \sum_{\spin_{l-1}} ... \Bigl( \sum_{\spin_{-l+1}} \Bigl( \sum_{\spin_{-l}}
P(\spin_{-l}|\alpha) \Bigr) P(\beta\spin_{-l+1}|\spin_{-l}\alpha)\Bigr)... P(\beta\spin_l|\spin_{l-1} ... \spin_{-l}\alpha) \Bigr) P(\beta|\spin_l \spin_{l-1} ... \spin_{-l} \alpha),
\ee
where 
\be
P(\spin_i|\spin_{i-1}\spin_{i_2}...\spin_{-l}\alpha)=
\frac{
\langle \psi^{mps}(t_{init})| 
\Pi^{i}_{\spin_i}
\Pi^{i-1}_{\spin_{i-1}}
\Pi^{i-2}_{\spin_{i-2}}
...
\Pi^{-l}_{\spin_{-l}}
\Pi^L_{\alpha}
| \psi^{mps}(t_{init})\rangle}
{
\langle \psi^{mps}(t_{init})| 
\Pi^{i-1}_{\spin_{i-1}}
\Pi^{i-2}_{\spin_{i-2}}
...
\Pi^{-l}_{\spin}
\Pi^L_{\alpha}
| \psi^{mps}(t_{init})\rangle}.
\ee

Thus, we can first sample $\alpha$ randomly as before, and then randomly sample the spin
on site $-l$ conditioned on that $\alpha$, randomly sample the spin on site $-l+1$ conditioned
on the given $\alpha$ and $\spin_{-l}$, and so on, until all spins are sampled, and then finally
sample $\beta$ according to
\be
P(\beta|\spin_l...\spin_{-l}\alpha)=
\frac{
\langle \psi^{mps}(t_{init})| 
\Pi^R_{\beta}
\Pi^{l}_{\spin_l}
...
\Pi^{-l}_{\spin_{-l}}
\Pi^L_{\alpha}
| \psi^{mps}(t_{init})\rangle}
{
\langle \psi^{mps}(t_{init})| 
\Pi^{l}_{\spin_{l}}
...
\Pi^{-l}_{\spin_{-l}}
\Pi^L_{\alpha}
| \psi^{mps}(t_{init})\rangle}.
\ee

Computing each of these conditional probabilities can be done in a time ${\cal O}(k_{max}^2)$
as follows.  After $j$ steps, we store the vector
$A^{A,B}(\spin_{-l+j-1}) ...
A^{A,B}(\spin_{-l})
\Pi^L_{\alpha}|\psi^{mps}(t_{init})\rangle$, where the $A$ matrices in this expression
are alternately chosen to be $A^A$ or $A^B$ depending on whether the site is on
the odd or even sublattice.
We then compute the two different vectors
$A^{A,B}(\spin_{-l+j})
A^{A,B}(\spin_{-l+j-1}) ...
A^{A,B}(\spin_{-l})
\Pi^L_{\alpha}|\psi^{mps}(t_{init})\rangle$
for the two different choices of $\spin_{-l+j}$, and compute the norms of these vectors to
determine the conditional probability.  After choosing $\spin_{-l+j}$, we keep the corresponding vector
and use it in the next step.  The matrix-vector multiplication takes a time ${\cal O}(k_{max}^2)$.

This solves the problem of statistically sampling $\alpha,\beta$.
Finally, we would like to discuss why a different, seemingly more
natural approach to the problem is not as good as the approach explained
here.  This different approach would be first statistically sample $\alpha$
as discussed here.  Then, initialize a matrix $\rho$ to the state
$|\alpha\rangle\langle\alpha|$.  The matrix product state defines a CP
map.  In fact, it defines two different CP maps, one for odd sites and one
for even sites, which we can call ${\cal E}^{even},{\cal E}^{odd}$.
Propagate the state $\rho$ through these two CP maps,
alternately applying the even and odd CP maps,
applying a total of $2l+1$ maps.  If $l$ is even, the result is
${\cal E}^{even}({\cal E}^{odd}...({\cal E}^{even}(\rho)...)$, while
if $l$ is odd then instead we apply
${\cal E}^{odd}({\cal E}^{even}...({\cal E}^{odd}(\rho)...)$.
The result gives the reduced
density matrix on sites $-\infty,...,+l$, given that on sites $-\infty,...,-l$
the system is in state $\alpha$.  The diagonal entries of this density
matrix give the probability distribution of $\beta$.  However, this approach, while
more natural, requires performing matrix-matrix multiplications, and hence
would be much slower than the matrix-vector multiplications that we used above.

It is important to understand that after sampling $\alpha,\beta$, we do not make
any further use of the randomly sampled values of
the spins $\spin_{-l},...,\spin_l$ that we found in this step.
We discard those values $\spin_{-l},...,\spin_l$ and keep just the $\alpha,\beta$
for the next step.
Different spin configurations associated with a given $\alpha,\beta$
will be added coherently, as seen in the next step where we compute
the amplitude to be in each spin configuration for the given $\alpha,\beta$.

\subsection{Computing $\Psi^{mps}_{\alpha,\beta}$}
The next step is to compute the amplitudes for $\Psi^{mps}_{\alpha,\beta}$ in the computational
basis.  There are $2^{2l+1}$ different amplitudes that we need to compute.
We first explain a simple (and slow) way of computing these amplitudes, and then
describe a much faster ``meet-in-the-middle" way to compute these amplitudes which is what
we actually used.

The simple way is as follows.  Suppose $l$ is even (if $l$ is odd, then the
sublattice indices $A,B$ will be reversed in this paragraph and the next).  We compute the amplitude
to be in each of the $2^{2l+1}$ different states in turn for the state
$\Pi^L_{\alpha}\Pi^R_{\beta} |\psi^{mps}(t_{init})\rangle$, and then we normalize the state after.
To compute the {\it unnormalized}
amplitude to be in the state $\spin_l,\spin_{l-1},...,\spin_{-l}$, first
compute
the inner product $\langle\alpha|A^A(\spin_{-l}) ... A^{B}(\spin_{l-1})
A^A(\spin_l)|\beta\rangle$, where
$\langle \alpha|$ is the vector with a $1$ in the $\alpha$-th entry, and
zeroes elsewhere.  Computing the inner product takes a time $(2l+1) {\cal O}(k_{max}^2)$.
Thus, the total time is
\be
(2l+1) 2^{2l+1} {\cal O}(k_{max}^2).
\ee
Note that since we compute unnormalized amplitudes, it does not matter that
the inner product
$\langle\alpha|A^A(\spin_{-l}) ... A^{B}(\spin_{l-1})
A^A(\spin_l)|\beta\rangle$ has an extra factor of $\lambda^A$ at the end compared to the
expression
$\langle\alpha|\Gamma^A(\spin_{-l}) \lambda^A ... \lambda ^A A^{B}(\spin_{l-1}) \lambda^B
\Gamma^A(\spin_l)|\beta\rangle$.  We will normalize the amplitudes later.

A much faster way is a meet-in-the-middle approach.  For each of the $2^{l+1}$ different
configurations of the spins $\spin_{-l},...,\spin_{-1},\spin_{0}$, we compute the
vector
$\langle \alpha|A^A(\spin_{-l})  ...A^{A}(\spin_{0})$.
Note that this vector is in the auxiliary Hilbert space of dimension given by the bond
dimension.
Also, for each of the $2^l$ configurations
of the spins $\spin_l,\spin_{l-1},...,\spin_{1}$, we compute the
vector
$A^B(\spin_{1}) ... A^{A}(\spin_{l})|\beta\rangle$.
Computing all of these vectors takes a time
\be
(l+1) 2^{l+1} {\cal O}(k_{max}^2).
\ee
We save all of these vectors (the memory requirement for this is ${\cal O}(k_{max} 2^{l+1})$ which
means that as long as $2^{l+1}\leq k_{max}$ the additional memory required is negligible).
Then, for each of the $2^{2l+1}$ spin configurations of spins $\spin_{l},...,\spin_{-l}$
we compute the inner product of the vectors
$\langle \alpha|A^A(\spin_{-l}) ... A^{A}(\spin_{0})$
and $A^B(\spin_{1})  ... A^{A}(\spin_{l}|\beta\rangle$.
This takes a time
\be
2^{2l+1} {\cal O}(k_{max}),
\ee
and hence the total time required is
\be
(l+1) 2^{l+1} {\cal O}(k_{max}^2)+
2^{2l+1} {\cal O}(k_{max}).
\ee

Again it is possible to make use of symmetry in these calculations.
We compute the vector
$A^B(\spin_{1}) ... A^{A}(\spin_{l})|\beta\rangle$ in $l$ different multiplications, and
after each multiplication the vector has a definite $S^z$.  This greatly speeds up the matrix-vector multiplications.
Also, the resulting vector
$\Psi^{mps}_{\alpha,\beta}$ has a non-vanishing amplitude only in a single
$S^z$ sector, which reduces the memory requirement and speeds up the calculation of the time evolution
described in the next subsection.

\subsection{Time Evolution of $\Psi^{mps}_{\alpha,\beta}$}
Finally, we time evolve the state $\Psi^{mps}_{\alpha,\beta}$ for time $t_{fin}-t_{init}$.  We
do this evolution using sparse matrix-vector multiplication as follows.
Before doing any Monte Carlo sampling of the states, we build a representation of the
Hamiltonian as a sparse matrix.  Each row of the matrix has ${\cal O}(l)$ non-vanishing elements.

Then, we fix a small time step $\delta t$ and evolve the vector over several of
these small time steps.  In the simulations here, we choose $\delta t=1/3$.  In this way,
a single Monte Carlo sample allows us to compute the expectation value of $S^z_0$ at several different
times.
To evolve for a time step $\delta t$ we approximate
\be
\exp(i H \delta t)|\Psi\rangle\approx \sum_{n=0}^{n_{max}} (-i)^n \frac{H^n}{n!}|\Psi\rangle,
\ee
where we truncate the Taylor series expansion at a finite order $n_{max}$.  The
vector
$H^n|\Psi\rangle$ can be computed using sparse matrix-vector multiplication.
The total computational time required is
\be
{\cal O}\Bigl(n_{max} (l+1) 2^{2l+1}\Bigr).
\ee
In practice, this step is very fast compared to the previous step of computing
$\Psi^{mps}_{\alpha,\beta}$ for the values of $k_{max}$ and $l$ that we choose.

\subsection{Pseudo-Code}

We recap this description of the procedure by giving the pseudo-code for a single Monte Carlo step.
\begin{itemize}
\item[{\bf 1.}] Choose $\alpha$ randomly according to the probability distribution $|\lambda^B(\alpha)|^2$.

\item[{\bf 2.}] {\bf For} $i=-l$ to $i=+l$ {\bf do}
\begin{itemize}
\item[{\bf 2a.}] 
Use the stored vector 
$A^{A,B}(\spin_{i-1})
...
A^{A,B}(\spin_{-l})
\Pi^L_\alpha
|\psi^{mps}(t_{init})\rangle$
to compute 
$
A^{A,B}(\spin_{i})
A^{A,B}(\spin_{i-1})
...
A^{A,B}(\spin_{-l})
\Pi^L_\alpha
|\psi^{mps}(t_{init})\rangle$
 for both choices of $\spin_{-l+j}$.

\item[{\bf 2b.}] Randomly choose $\spin_{-l+j}$ according to
$P(\spin_i|\spin_{i-1}\spin_{i-2}...\spin_{-l}\alpha)$.
\end{itemize}

\item[{\bf 3.}] Randomly sample $\beta$ according to
$P(\beta|\spin_l...\spin_{-l}\alpha)$.

\item[{\bf 4.}] For each of the $2^{l+1}$ configurations of spins $-l,...,0$ compute the vector
$\langle \alpha|A^A(\spin_{-l}) ... A^A(\spin_{0})$.  Save for use in step {\bf 6}.

\item[{\bf 5.}]
For each of the $2^l$ configurations
of the spins $\spin_l,\spin_{l-1},...,\spin_{1}$, compute the
vector
$A^B(\spin_{1}) ... A^{B}(\spin_{l})|\beta\rangle$.  Save for use in step {\bf 6}.

\item[{\bf 6.}]
For each of the $2^{2l+1}$ spin configurations of spins $\spin_{l},...,\spin_{-l}$
compute the inner product of the vectors
$\langle \alpha|A^A(\spin_{-l}) ... A^{A}(\spin_{0})$
and $A^B(\spin_{1}) ... A^{A}(\spin_{l})|\beta\rangle$.
Save the result as the amplitudes of the state $\Psi^{mps}_{\alpha,\beta}$ in the computational basis.

\item[{\bf 7.}] Normalize that state $\Psi^{mps}_{\alpha,\beta}$.

\item[{\bf 8.}] Evolve state $\Psi^{mps}_{\alpha,\beta}$ forward by a sequence of time steps $\delta t$ until
time $t_{fin}$.  Compute spin expectation on site $0$ after each time step.
\end{itemize}

\subsection{Time Estimates}
The total time required for all steps is
\be
{\cal O}\Bigl(l 2^l k_{max}^2+l (t_{fin}/\delta t) n_{max} 2^{2l}\Bigr).
\ee
The dependence on $t_{fin}$ and $n_{max}$ is unimportant compared to the exponential dependence on $l$, since
it suffices to take $n_{max}={\cal O}(l)$ to obtain accurate results.  Looking only at exponential dependence
on $l$ and polynomial dependence on $k_{max}$, and ignoring any polynomial dependence on $l$, we find
\be
k_{max}^2 2^l+2^{2l}.
\ee
This is to be compared with a time ${\cal O}(k_{max}^3)$ for the matrix product simulation (again, ignoring the
dependence on the number of Trotter-Suzuki states, and only considering the dependence on $k_{max}$).
Thus, equating the two times, we find that we can take
\be
2^l\sim k_{max}.
\ee
This in fact is roughly the regime in which we worked below, since we took $l=10$ and $k_{max}=4096$.
Similarly, since we work in the regime $2^{l+1}<k_{max}$, our additional memory requirements are negligible.

In fact, our time to do a single Monte Carlo step was much faster than our time to do the matrix product
simulation.  This was balanced by the fact that we have to do many Monte Carlo steps.  Fortunately, the Monte
Carlo steps parallelize trivially.

In principle one could avoid any Monte Carlo sampling by summing over all possible values of $\alpha,\beta$,
appropriately weighted.  While this calculation parallelizes well, it would not be practical
for large values of $k_{max}$.

\section{Results}

We simulated using the modified iTEBD for $k_{max}$ ranging up to $4096$.
Accurate results were found up to a time slightly greater than $16$,
with low truncation error.
Unfortunately, we do not know a good way to directly compute truncation error in iTEBD.  For a finite
size system, one can compute the difference between the state after truncation and the state before truncation.
This gives an upper bound on how the truncation affects expectation values.  For an infinite system, this doesn't
work.  To see why this doesn't work, consider a finite but large system: even if we make only a small error in the
state everywhere and are able to accurately approximate all the local expectation values,
we make a large total error in the state vector because the system is large.  So, we estimated truncation error
in a few different ways.  We checked how certain observables, such as $S^z_0$, depended on $k_{max}$.  For
a given $k_{max}<4096$, the curve would follow the $k_{max}=4096$ curve for some time, and then deviate.
We could see how the time of deviation depended on $k_{max}$, and extrapolate to $k_{max}=4096$.

Also, we took results from the iTEBD simulation at times $15.0625$ and $16.0625$ and started the Monte Carlo sampling
at those two different times and compared results.  When we compared two different Monte Carlo averages,
one with $t_{init}=16.0625$ and one with $t_{init}=15.0625$, both having the same $l$, we observed that the
results agreed, up to statistical error,
until a time $t_{fin}\approx 15.0625+l/v_{sw}$.  That is, the deviation between the two
curves only arose because the curve started at $t_{init}=15.0625$ saw the finite-size effect of a finite $l$ earlier
than did the curve started at $t_{init}=15.0625$.  Thus, we are fairly confident about the accuracy of our
modified iTEBD simulation up to time $16.0625$ up to times of roughly
$22.5$ as shown in the figures below.

It took roughly 2 CPU days on a 2Ghz Opteron to do 1000 Monte Carlo samples for
$l=10$.
We did 98000 samples for $t_{init}=15.0625$ and $82000$ for $t_{init}=16.0625$.
We did several additional runs for smaller $l$ and smaller $t_{init}$ to
check the algorithm.
In Fig.~3 we show the computed behavior of the magnetization ($\langle
S^z_0(t) \rangle$) as a function of time using the modified iTEBD with
$k_{max}=1024,2048,4096$.  For any given $k_{max}$, the accuracy of the
algorithm breaks down badly at sufficiently long time.  Doubling $k_{max}$
leads to only a constant increase in the time which can be simulated, which
implies that reaching times $>20$ will require prohibitively large matrices.

\begin{figure}
\label{ifig}
\centerline{
\includegraphics[scale=0.4]{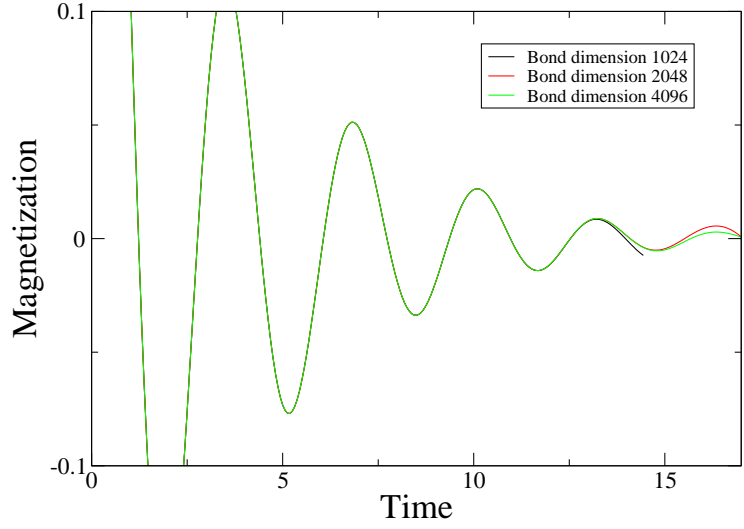}}
\caption{Comparison of iTEBD for $k_{max}=1024,2048,4096$.}
\vspace{5mm}
\end{figure}

In Fig.~4 we show a comparison of the light-cone method and iTEBD
with $k_{max}=4096$.  The difference between the two different light-cone
curves is within sampling error for the given number of samples.  The 
root-mean square sampling
error was roughly $0.00025$.  By increasing the number of samples this
error can be reduced as the square-root of the number of samples.

\begin{figure}
\label{dfig}
\centerline{
\includegraphics[scale=0.4]{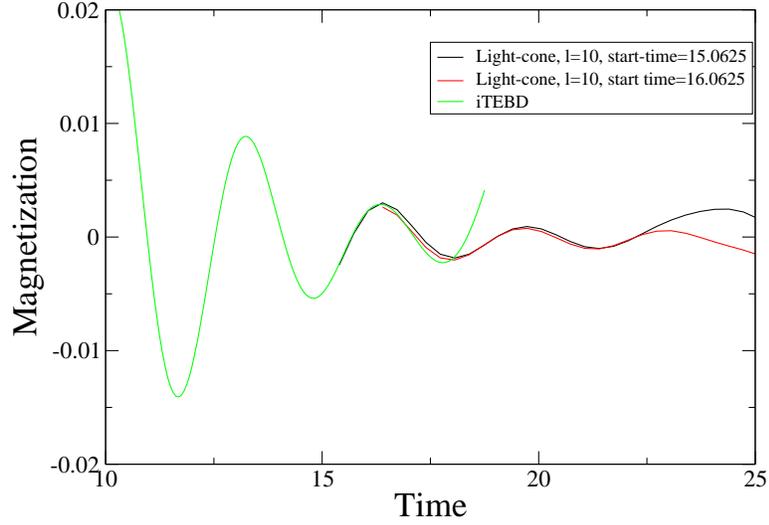}}
\caption{Comparison of light-cone and iTEBD with $k_{max}=4096$.}
\vspace{5mm}
\end{figure}

In Fig.~5 we show the mean-field of \cite{hl}.  This curve shows
an increase in the order parameter at times $>20$, followed by a decrease
at later times.  The curve is well-described by beating together two
different cosines, with a $1/t^{3/2}$ envelope.  The revivals come from the
beats.

\begin{figure}
\label{mffig}
\centerline{
\includegraphics[scale=0.4]{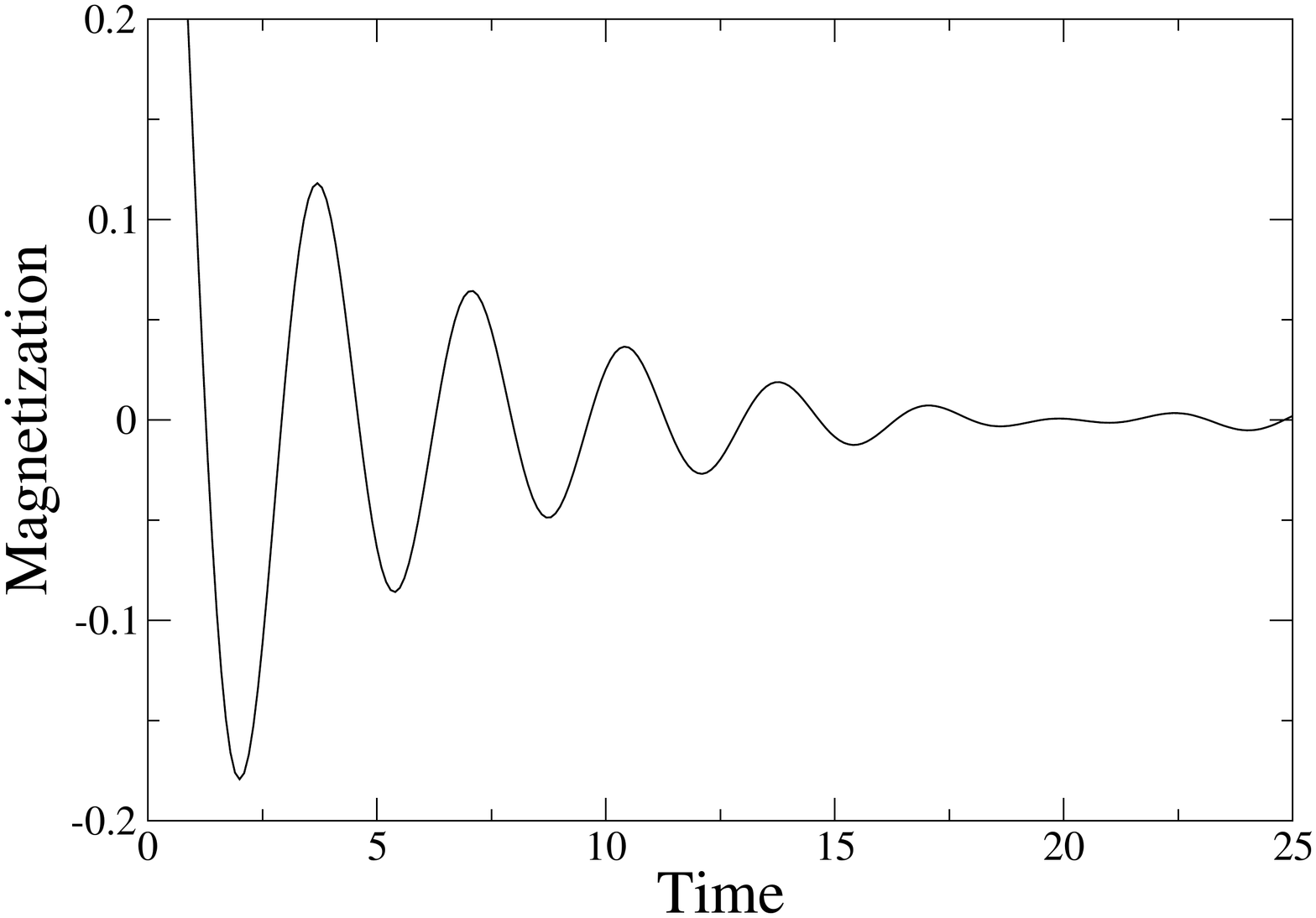}}
\caption{Mean-field of \cite{hl}.}
\vspace{5mm}
\end{figure}

In the black line in Fig.~6 we show the time series of peak heights for each
maximum in the absolute value of $\langle S^z_0(t) \rangle$.  Each circle
represents a given time at which a maximum occurred.  Circles at early times
are from iTEBD, circles at later times are from light-cone.  Note the
non-monotonic behavior.  The $y$-axis is a logarithmic scale.
The non-monotonic behavior is within sampling error,
so it is possible that the peaks do decay monotonically.  However, we
are fairly confident that even if the peaks do decay monotonically there
is a distinct feature in the curve showing a flattening of the decay
at later times.

\begin{figure}
\label{pkfig}
\centerline{
\includegraphics[scale=0.4]{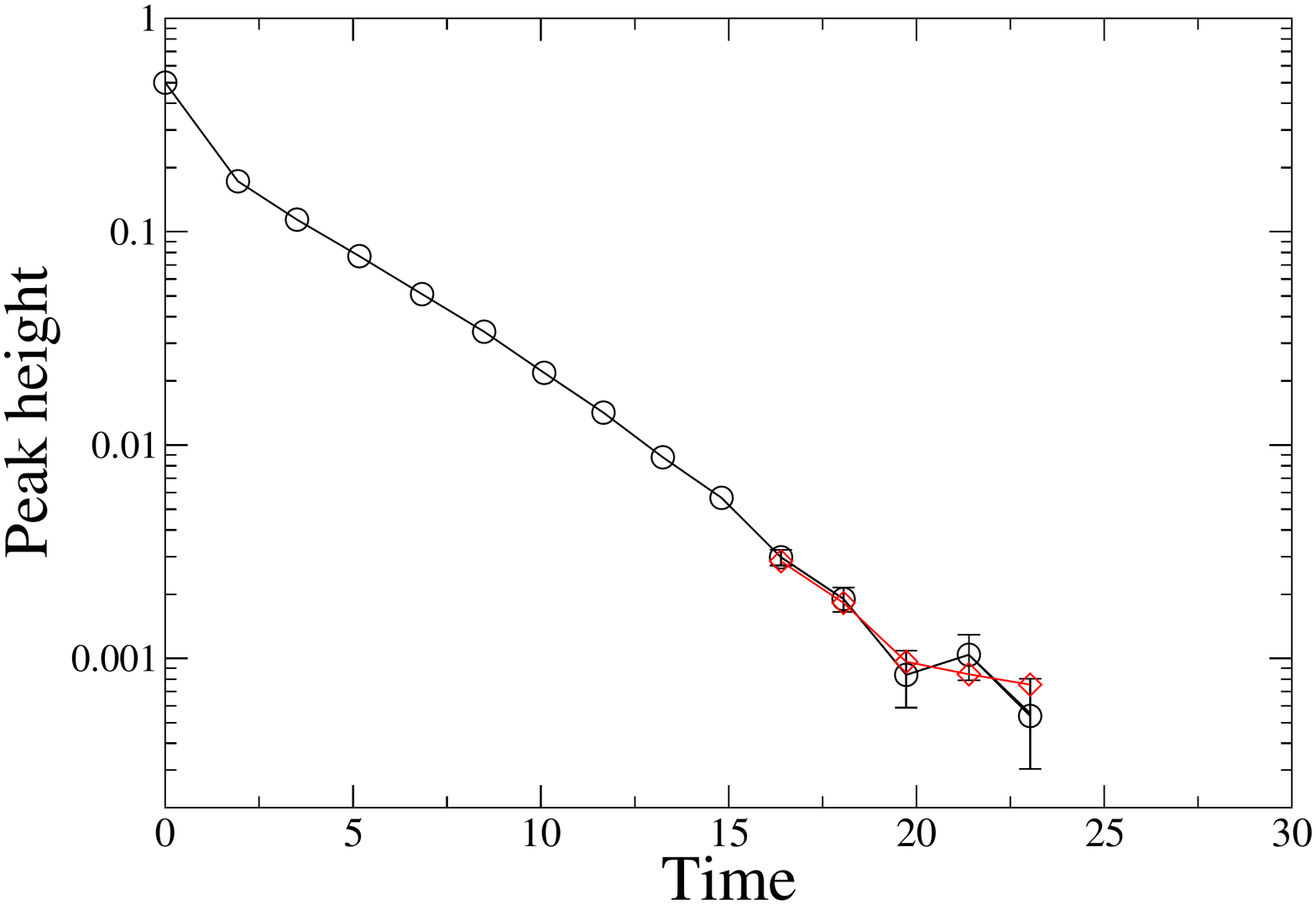}}
\caption{Peak heights.  Error bars represent sampling error.  Black line
is raw data.  Red line is with constant factor added to $y$-value of sampled
data to correct for greater variance in the mean as described in text.}
\vspace{5mm}
\end{figure}

It is possible to reduce the sampling error further by more runs.  A
ten-fold increase in runs would enable more definite statements about the
monotonicity of the peak heights.  Interestingly, the root-mean-square
fluctuation in the mean of a given light-cone curve (such as the curve
with $t_{init}=16.0625$) is much greater than the root-mean-square fluctuation
in the difference of the $y$-coordinate of the curve at two different
times.  So, it may be possible to reduce sampling error by shifting the
$y$-axis by a constant to agree with the known correct result at early
times.  We have tested this, as shown in the red line in Fig.~6
and the result indicates that the decay of peak heights almost stops
at around time $t=20$, although in this case the peak heights no longer
increase at time $21.5$.  More runs will be required to be certain 
whether the peak height is actually non-montonic or whether there is simply
a sharp shoulder in the graph of peak heights.

\section{Discussions}
We have developed a method combining the light-cone technique with matrix product methods.  While
the computational complexity is still exponentially large as a function of time, the exponent
is lower than using matrix product methods alone.  As a result, we are able to go to
significantly longer times for the same computational effort.  We have discussed several
different ways of doing the sampling depending on the computational resources of time and memory.

While we used a cluster to perform statistical averaging, the light-cone method makes it possible to
perform simulations for shorter times with very small computational cost.  If we reduce $t_{init}$,
then the required $k_{max}$ reduces also, and then the statistical sampling becomes faster (the dominant
cost in the statistical sampling is the ${\cal O}(2^{l} k_{max}^2)$ cost).  For example, with $k_{max}=128$, it is
possible to reach $t_{init}\approx 11$; in that case if we take the same $l=10$, we could reach a time
of roughly $18$.  With $k_{max}=128$, the statistical sampling would of course be much faster than with
$k_{max}=4096$.

Our results support the possibility of either revivals in the order
parameter or of a flattening out of the decay of peak height.  Perhaps at smaller $\Delta$ the system would be closer to mean-field
and so the revivals would be more clear; however, at the same time, the time
before revivals occur is longer in this case which makes the simulations
more difficult.

The integrability of the system does not play any role in our method.  However,
it does play some role in the results.  Even in the mean-field
studied in \cite{hl}, breaking integrability leads to a large change
in the asymptotic behavior of the magnetization; the mean-field for the
integrable system has a power law decay, while non-integrable perturbations
lead to a much faster exponential decay and can destroy the revivals if they
are strong enough.

{\it Acknowledgments---} 
I thank R. G. Melko for useful discussions on numerical simulation.  I thank L. Levitov for useful
discussions about mean-field behavior in these systems.
I thank the CNLS for use of the cnlsopts
cluster.  I thank the Rocky Mountain Bio Lab for hospitality.
This work was
supported by U. S. DOE Contract No. DE-AC52-06NA25396.

\end{document}